\documentclass[11pt]{article}

\usepackage{amsmath}
\usepackage{graphicx}
\usepackage{amsfonts}
\usepackage{amssymb}
\usepackage{epsfig}
\usepackage{color}
\usepackage{psfrag}

\newcommand{\EQ}{\begin{equation}}
\newcommand{\EN}{\end{equation}}
\newcommand{\be}{\begin{equation}}
\newcommand{\ee}{\end{equation}}
\newcommand{\bea}{\begin{eqnarray}}
\newcommand{\eea}{\end{eqnarray}}

\begin{document}

\setcounter{page}{0} \topmargin0pt \oddsidemargin5mm \renewcommand{%
\thefootnote}{\arabic{footnote}}\newpage \setcounter{page}{0} 
\begin{titlepage}
\begin{flushright}
SISSA 35/2010/EP
\end{flushright}
\vspace{0.5cm}
\begin{center}
{\large{\bf Universal properties of Ising clusters and droplets near criticality}
}\\

\vspace{1.8cm}
{\large Gesualdo Delfino
and Jacopo Viti
} 
\\
\vspace{0.5cm}
{\em International School for Advanced Studies (SISSA)}\\
{\em via Bonomea 265, 34136 Trieste, Italy}\\
{\em INFN sezione di Trieste}\\
\end{center}
\vspace{1.2cm}

\renewcommand{\thefootnote}{\arabic{footnote}}
\setcounter{footnote}{0}

\begin{abstract}
\noindent
Clusters and droplets of positive spins in the two-dimensional Ising model percolate at the Curie temperature in absence of external field. The percolative exponents coincide with the magnetic ones for droplets but not for clusters. We use integrable field theory to determine amplitude ratios which characterize the approach to criticality within these two universality classes of percolative critical behavior. 
\end{abstract}

\end{titlepage}

\newpage

\section{Introduction}
The simplest observable one can think of within the lattice modellization of a ferromagnet is the average value of the spin at a given site. On an infinite regular lattice this gives the magnetization per site $M$, which serves as order parameter of the ferromagnetic transition. It is also natural, however, to look at extended (non-local) objects like the clusters formed by neighboring spins with the same value. Then the probability $P$ that a given site belongs to an infinite cluster provides the order parameter of a percolative phase transition. The relation between the magnetic and percolative transitions within the ferromagnet is far from trivial and has been the subject of many studies \cite{SA}, first of all for the basic case of the Ising ferromagnet which is also the subject of this paper. 

The coincidence of the two transitions (concerning both the location of the critical point and the critical exponents) was the requirement of the droplet model for ferromagnetism \cite{Fisherdroplet}. It is not fulfilled by the ordinary clusters defined above, but is satisfied by special clusters whose mass is suitably reduced as the temperature increases \cite{CK}; these particular clusters are then called droplets. On usual lattices in two-dimensions also the ordinary clusters percolate at the Curie temperature $T_c$ at which the magnetic transition takes place \cite{CNPR}, although in this case the percolative and magnetic critical exponents do not coincide \cite{SG}. As a result, $T_c$ is simultaneaously the location of the ferromagnetic transition and of the percolative transition for both clusters and droplets. 

A formulation of the problem suitable for theoretical study, for both clusters  and droplets, is obtained coupling the Ising spins to auxiliary color variables whose expectation value becomes the percolative order parameter $P$ \cite{KF,Murata}. In this way also the cluster properties are related to correlation functions of local variables and can be studied using the renormalization group \cite{CK} and field theory \cite{CL,BC}. In particular, the exact results of two-dimensional conformal field theory \cite{BPZ} allowed the identification of the critical exponents also for clusters \cite{SV}.

In this respect, it is worth recalling that in recent years the role of non-local observables within spin models has been much emphasized in connection with Schramm-Loewner evolution (SLE) (see e.g. \cite{CardySLE,BB} for reviews). Indeed, the latter provides an approach to two-dimensional critical behavior based on the study of conformally invariant random curves which may be thought as cluster boundaries. Some exponents and other critical properties have been derived in this way within an approach alternative to conformal field theory. On the other hand, moving away from the critical point in the SLE framework still appears a difficult task, and very few steps have been done in this sense (see e.g. \cite{BBC,MS}). 

Moving away from criticality within field theory is, on the contrary, very natural and, in many cases, can be done preserving integrability \cite{Taniguchi}. It was shown in \cite{isingperc} how Ising clusters and droplets near criticality can be described using perturbed conformal field theory and, in particular, how the second order transition that clusters undergo above $T_c$ in an external field becomes integrable in the scaling limit. In this paper we use the field theoretical setting of \cite{isingperc} to quantitatively characterize the universal properties of clusters and droplets in their approach to criticality. This is done exploiting integrability to compute universal combinations of critical amplitudes of the main percolative observables, namely the order parameter, the connectivity length, the mean cluster number and the mean cluster size.

We obtain new results for clusters, but also for droplets. This may appear surprising in consideration of the fact that droplets provide the percolative description of the magnetic transition, for which all canonical universal ratios are known exactly \cite{WMcTB,report}. The point is that, while the critical exponents are determined by the singularities of correlation functions at distances much shorter than the correlation length $\xi$ (in the scaling limit $T\to T_c$, where $\xi$ is anyway very large), the amplitude ratios are also sensitive to correlations at larger distances. The spin-spin correlator and the connectivity within finite droplets have the same singular behavior at short distances but, due to the contribution of infinite droplets, differ at larger distances below $T_c$. As a result, magnetic and percolative exponents coincide, while some amplitude ratios differ. 

A characterization of cluster properties similar to that of this paper was done in \cite{DC,DVC} for the case of random percolation. Of course the important physical difference with respect to that case is that in Ising percolation cluster criticality is determined by the ferromagnetic interaction. The site occupation probability is not an independent parameter to be tuned towards its critical value, as in the random case, but is instead a function of temperature and magnetic field. One visible manifestation of this difference within the formalism arise in the evaluation of correlation functions through the spectral decomposition over intermediate particle states. While in random percolation all degrees of freedom, and then all particles, are auxiliary, in Ising percolation the particles associated to the magnetic degrees of freedom are also part of the game. This leads us to formulate selection rules which identify the particle states actually contributing to the percolative properties. The picture which emerges is that of a sharp separation between states contributing to magnetic correlations and states contributing to cluster connectivity. For the mean cluster number, which is related to the free energy, the presence of the magnetic interaction results into the appearance of logarithmic terms which are absent in the random case. Altogether, our analysis produces a number of universal field theoretical predictions for both clusters and droplets, and for different directions in parameter space, that will be interesting to compare with lattice estimates when these will become available.

The paper is organized as follows. In the next section we recall how percolative observables are described in terms of auxiliary color variables before turning to the characterization of their behavior near criticality in section~3. Section~4 is devoted to the field theory description, while universal amplitude combinations are discussed in section~5. Few conclusive remarks and two appendices complete the paper.

\section{Kasteleyn-Fortuin representation}
We consider the ferromagnetic Ising model defined by the reduced 
Hamiltonian
\EQ
-{\cal H}_{\text{Ising}}=\frac1T\sum_{<x,y>}\sigma(x)\sigma(y)+H\sum_x
\sigma(x)\,,
\label{ising}
\EN
where $\sigma(x)=\pm 1$ is a spin variable located at the site $x$ of an infinite 
regular lattice $\mathbb L$, $T\geq 0$ and $H$ are couplings that we call temperature and magnetic field, respectively, and the first sum is restricted to 
nearest-neighbor spins.

Correlated percolation in the Ising model can be conveniently studied by
coupling Ising spins to auxiliary Potts variables taking the values (colors) $s(x)=1,\ldots,q$. 
Replacing Ising spins with lattice gas variables $t(x)=\frac{1}{2}(\sigma(x)+1)=0,1$, the resulting model is a ferromagnetic dilute q-state Potts model with
Hamiltonian  
\EQ
-\mathcal{H}_q=\frac{4}{T}\sum_{<x,y>}t(x)t(y)+\Delta\sum_x
t(x)+J\sum_{<x,y>}t(x)t(y)\left(\delta_{s(x),s(y)}-1\right)+\tilde{h}\sum_x\left(\delta_{s(x),1}-\frac{1}{q}\right)t(x),
\label{hamiltonian}
\EN
where $\Delta=2H-a/T$, with $a$ a lattice-dependent constant, and we allow for the presence of a field $\tilde{h}$ which explicitly breaks the $S_q$ invariance under permutations of the $q$ colors. The Potts spins
\EQ
\sigma_k(x)\equiv\left(\delta_{s(x),k}-\frac{1}{q}\right)t(x),\hspace{1cm} k=1,\ldots,q
\label{pottspin},
\EN
effectively live only on the restricted lattice $\mathbb L_0$ formed by the sites of $\mathbb L$ with positive Ising spin (i.e. with $t(x)=1$). 

The partition function $Z_q$ associated to the Hamiltonian (\ref{hamiltonian}) admits the following Kasteleyn-Fortuin (KF) representation \cite{KF,Murata}
\bea
 Z_q & =&  \sum_{\{t(x)\}}\sum_{\{s(x)\}}\text{e}^{-\mathcal{H}_q}\nonumber\\
 & =& \sum_{\mathbb L_0\subseteq\mathbb
  L}\text{e}^{-\mathcal{H_{\text{Ising}}}}q^{N_e}\sum_{\mathcal{G}}\Lambda(\mathcal{G})\prod_c\left[\text{e}^{\tilde{h}\left(1-\frac{1}{q}\right)s_c}+(q-1)\text{e}^{-\tilde{h}\frac{s_c}{q}}\right],
\label{KF}
\eea
where $N_e$ is the number of ``empty'' sites (i.e. having $t(x)=0$ and then not belonging to $\mathbb L_0$). For any lattice gas configuration $\{t(x)\}$, the sum over the Potts variables $s(x)$ is transformed into a sum
over all possible graphs $\mathcal{G}$ obtained drawing bonds between nearest neighbors belonging to $\mathbb L_0$. The weight associated to each such a bond 
is $p_B=1-\text{e}^{-J}$ (bond occupation probability), so that the weight of a graph ${\cal G}$ is $\Lambda(\mathcal{G})=p_B^{n_B(\mathcal{G})}(1-p_B)^{\overline{n_B}(\mathcal{G})}$, where $n_B(\mathcal{G})$ ($\overline{n_B}(\mathcal{G})$) is the number of bonds occupied (unoccupied) on $\mathbb L_0$. Connected components of $\mathcal{G}$ are called KF clusters and $s_c$ is the number of sites in the $c$-th cluster\footnote{Unconnected sites on $\mathbb L_0$ also counts as clusters with $s_c=1$.}. 

For $\tilde{h}=0$ the product over clusters in (\ref{KF}) reduces to $q^{N_c}$, $N_c$ being the number of KF clusters in ${\cal G}$. The factor $q^{N_e+N_c}$ disappears in the limit $q\to 1$, so that (\ref{KF}) defines a percolative average for KF clusters living on Ising clusters. In particular, the KF clusters become the Ising clusters when $p_B=1$, i.e. when $J\to+\infty$. In this section we refer to the general case of KF clusters.

Standard definitions for percolative observables apply. The percolative order parameter $P$ is the probability that the site in the origin belongs to an infinite cluster, namely the average fraction of sites of $\mathbb L$ belonging to infinite clusters.
The average size of finite clusters containing the origin is 
\EQ
S=\frac{1}{N}\langle\sum_{c}{}' s_{c}^2\rangle\,,
\label{size}
\EN
where the primed sum runs over finite clusters only and $N$ is the number of sites in $\mathbb L$, which diverges in the thermodynamic limit we are considering. The probability $P_{f}(x)$ that the origin and $x$ belong to the same finite cluster defines the `true' and `second moment' connectivity lenghts $\xi_t$ and $\xi_\text{2nd}$ through the relations
\bea
&& P_{f}(x)\sim\text{e}^{-|x|/\xi_t},\quad|x|\rightarrow\infty\,,
\label{true}\\
&& \xi_\text{2nd}^2=\frac{\sum_x|x|^2P_f(x)}{4\sum_xP_f(x)}\,.
\label{2nd}
\eea
The observables $P$ and $S$ are related to the dilute Potts magnetization and susceptibility, respectively. Indeed the Potts spontaneous magnetization is 
\EQ
\langle\sigma_1(x)\rangle=\frac{1}{N}\left.\frac{\partial\ln
  Z_q}{\partial\tilde{h}}\right|_{\tilde{h}=0^+}=\lim_{\tilde{h}\rightarrow 0^+}\langle\sum_c F(s_c, \tilde{h},
q)\rangle,
\label{1pt}
\EN
where 
\EQ 
F(s_c, \tilde{h},
q)=\left(1-\frac{1}{q}\right)\frac{s_c}{N}\frac{\text{e}^{\tilde{h}\left(1-\frac{1}{q}\right)s_c}-\text{e}^{-\frac{\tilde{h}}{q}s_c}}{\text{e}^{\tilde{h}\left(1-\frac{1}{q}\right)s_c}+\text{e}^{-\frac{\tilde{h}}{q}s_c}(q-1)}.
\label{Ffunction} 
\EN
In the limit $\tilde{h}\rightarrow 0^+$, (\ref{Ffunction}) vanishes for any finite cluster, so that only infinite clusters contribute to $\sum_cF$ a term $(1-1/q)$ times the fraction of the lattice they occupy; hence we have the relation
\EQ
P=\lim_{q\to 1}\frac{\langle \sigma_1(x)\rangle}{q-1}\,,
\label{mag}
\EN
showing that the percolative transition of clusters maps onto the spontaneous breaking of $S_q$ symmetry in the auxiliary Potts variables. The Potts longitudinal susceptibility $\left.\frac{\partial^2\ln Z_q}{N\partial\tilde{h}^2}\right|_{\tilde{h}=0^+}$ can be expressed through the spin-spin correlator or differentiating the cluster expansion (\ref{KF}); this leads to the relation\footnote{Throughout the paper we denote connected correlators attaching a subscript $c$ to the average symbol.}
\EQ
S=\lim_{q\to 1}\frac{1}{q-1}\sum_{x}\langle\sigma_1(x)\sigma_1(0)\rangle_c\,.
\label{suscep}
\EN       

It also follows from (\ref{KF}) that the mean cluster number per site is given by
\bea
\frac{\langle N_c\rangle}{N}&=&\frac{1}{N}\left(\lim_{q\to 1}\partial_q\left.\ln Z_q\right|_{\tilde{h}=0}-\langle N_e\rangle\right)\nonumber\\
&=&-\left.\partial_qf_q\right|_{q=1}-\frac12(1-M)\,,
\label{mcn}
\eea
where $f_q=-(1/N)\left.\ln Z_q\right|_{\tilde{h}=0}$ is the dilute Potts free energy per site and $M$ is the Ising magnetization per site.

We conclude this section observing that (\ref{mag}) and (\ref{suscep}) can also be derived as follows. Take $\tilde{h}=0^+$ and denote by $\nu$ the average fraction of sites belonging to $\mathbb L_0$, and by $\nu_f$ the fraction of sites belonging to finite clusters. We have
\bea
& & \langle t(x)\rangle=\nu=P+\nu_f, \label{t}\\
& & \langle\delta_{s(x),1}t(x)\rangle=P+\frac{1}{q}\nu_f \label{delta}\,,
\eea
where we use the fact that a site has color $1$ with probability $1/q$ if it belongs to a finite cluster, and with probability $1$ if it belongs to an infinite cluster\footnote{Infinite clusters contribute only to the first term inside the product in (\ref{KF}) for $\tilde{h}\to 0^+$. In turn, only sites with color $1$ contribute to this term.}; equation (\ref{mag}) then follows recalling (\ref{pottspin}). Considering instead two sites $x$ and $y$, we call $P_{\alpha\beta}(x-y)$ the probability that $x$ is of type $\alpha$ and $y$ of type $\beta$, with the following specifications: $\alpha$ and $\beta$ take the value $f$ if the corresponding site belongs to a finite cluster, $i$ if it belongs to an infinite cluster, and $e$ if it does not belong to a cluster (i.e. it is empty); more precisely, $P_{ff}$ is the probability that the two sites belong to different finite clusters, while we call $P_f$ the probability that they belong to the same finite cluster and $P_i$ the probability that they both belong to infinite clusters. Introducing also the probability $P_{oo}$ that the sites both belong to some cluster, we can write the relations
\bea
& & P_{oo}+P_{ie}+P_{fe}=\nu, \\
& & P_{oo}+2(P_{ie}+P_{fe})+P_{ee}=1, \label{occ-empty}\\
& & P_{i}+2P_{if}+P_{f}+P_{ff}=P_{oo}, \label{occ}\\
& & P_{i}+P_{if}+P_{ie}=P.\label{occ-inf}
\eea
These leave four independent two-point probabilities that we choose to be
$P_{i}, P_{if}, P_{f}$ and $P_{ff}$. Through them we can express the four independent two-point spin correlators in the dilute Potts model as
\bea
& & \langle t(x)t(0)\rangle=P_{i}+2P_{if}+P_{f}+P_{ff},\label{tt}\\
& & \langle
\delta_{s(x),1}t(x)\,\,t(0)\rangle=P_i+\left(1+\frac{1}{q}\right)P_{if}+\frac{1}{q}\left(P_f+P_{ff}\right),
\label{deltat}\\
& & \langle
\delta_{s(x),1}t(x)\,\,\delta_{s(0),1}t(0)\rangle=P_i+\frac{2}{q}P_{if}+\frac{1}{q}P_{f}+\frac{1}{q^2}P_{ff},\label{long}\\
& & \langle\delta_{s(x),k}t(x)\,\,\delta_{s(0),k}t(0)\rangle=\frac{1}{q}P_{f}+\frac{1}{q^2}P_{ff},\hspace{1cm}k\neq 1.\label{tr}
\eea
These equations give in particular
\bea
G(x)\equiv\langle\sigma_1(x)\sigma_1(0)\rangle_c &=&\left(1-\frac{1}{q}\right)^2(P_i-P^2)+\frac{1}{q}\left(1-\frac{1}{q}\right)P_{f},
\label{connectivity}\\
\langle\sigma_k(x)\sigma_k(0)\rangle_c &=&\frac{1}{q^2}[(P_i-P^2)+(q-1)P_{f}],\hspace{1cm}k\neq 1,
\label{transverse}
\eea
and (\ref{suscep}) follows from the fact that $S=\lim_{q\to 1}\sum_x{P_{f}(x)}$.

\section{Clusters and droplets near criticality}
\begin{figure}
\centerline{
\includegraphics[width=7cm]{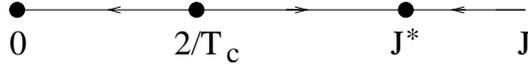}}
\caption{\label{jflows} Renormalization group flows in the coupling $J$ for the Hamiltonian (\ref{hamiltonian}) with $q\to 1$, $T=T_c$, $H=0$. While $J=0$ is just the Ising magnetic fixed point, $2/T_c$ and $J^*$ are percolation fixed points for Ising droplets and Ising clusters, respectively.
}
\end{figure}

The critical properties of KF clusters within the Ising model are ruled by the renormalization group fixed points of the dilute Potts Hamiltonian (\ref{hamiltonian}) with $\tilde{h}=0$ and $q\to 1$. Since the percolative properties do not affect the magnetic ones, we need to be at the magnetic fixed point $(T,H)=(T_c,0)$ to start with, and are left with the problem of finding fixed points of the coupling $J$. It was first argued in \cite{CK} that for the case $J>0$ of interest here there are two such fixed points, with a renormalization group pattern shown in Fig.~\ref{jflows}. The critical properties of the Ising clusters (the KF clusters with $p_B=1$, i.e. $J=+\infty$) renormalize onto those of the fixed point with the larger value of $J$, that we call $J^*$. This corresponds to the tricritical point of the dilute Potts model with $q\to 1$. 

The second fixed point, located at $J=2/T_c$, follows from the identity \cite{CK}
\EQ
\left.-{\cal H}_q\right|_{\tilde{h}=0,\,\,J=2/T}=\frac2T\sum_{<x,y>}(\delta_{\nu(x),\nu(y)}-1)+(\ln q-2H)\sum_x\delta_{\nu(x),0}\,,\hspace{.6cm}\nu(x)=0,1,\ldots,q,
\label{q+1}
\EN
showing that for $J=2/T$ the Ising and color variables can be combined into a single $(q+1)$-state Potts variable $\nu(x)$ taking the value $0$ on sites with negative Ising spin, and the values $1,\dots,q$ on sites with positive Ising spin; for $2H=\ln q$, (\ref{q+1}) exhibits a $S_{q+1}$ invariance whose spontaneous breaking yields the fixed point at $T=T_c$, $H=0$, in the limit $q\to 1$. It is natural to associate to the $(q+1)$-state Potts model (\ref{q+1}) the spin variables
\EQ
\omega_{\alpha}(x)=\delta_{\nu(x),\alpha}-\frac{1}{q+1},\hspace{.6cm}\alpha=0,1,\ldots,q,
\label{omega}
\EN 
whose average provides the order parameter of the phase transition. Using $\nu(x)=t(x)s(x)$ and $\sum_{\alpha=0}^q\omega_\alpha=\sum_{k=1}^q\sigma_k=0$, it is easy to check that the Ising site variable and the dilute Potts spin (\ref{pottspin}) can be written as
\bea
t(x)&=&-\omega_0(x)+\frac{q}{q+1}\,,
\label{omega0}\\
\sigma_k(x)&=&\omega_k(x)+\frac{\omega_0(x)}{q},\hspace{.6cm}k=1,\ldots,q\,.
\label{matching}
\eea
Denoting $\langle\cdots\rangle_\beta$ the average in the phase where the spontaneous breaking of $S_{q+1}$ permutational symmetry selects the direction $\beta$, we have
\EQ
\langle\omega_\alpha\rangle_\beta=[(q+1)\delta_{\alpha\beta}-1]\frac{M_{q+1}}{q}\,,
\label{vev}
\EN
and, using (\ref{matching}),
\EQ
\langle\sigma_1\rangle_\beta=\left\{
\begin{array}{l}
0\,,\hspace{2cm}\beta=0\\ 
\frac{q^2-1}{q^2}M_{q+1}\,,\hspace{.6cm}\beta=1\,.
\end{array}
\right.  
\label{sigma1beta}
\EN
At this point (\ref{mag}) implies $P=0$ in the phase $\beta=0$, and $P=M$ in the phase $\beta=1$, where $M=2M_2$ is the Ising spontaneous magnetization, as implied by (\ref{omega0}) and (\ref{vev}). This amounts to say that there is a first order percolative transition along the segment $T<T_c$, $H=0$, where the limit $H\to 0^\mp$ is described by the phases $\beta=0,1$, respectively, of (\ref{q+1}) with $q\to 1$. Hence, for $H=0$, the magnetic and percolative transitions have the same nature and location, and, due to the identity $P=M$ at $H=0^+$, the same critical exponents. Since these are the requirements of the droplet model \cite{Fisherdroplet} aimed at describing the magnetic transition as a percolation transition, the KF clusters with $p_B=1-e^{-2/T}$ are called {\em droplets}.

\begin{figure}
\centerline{
\includegraphics[width=7cm]{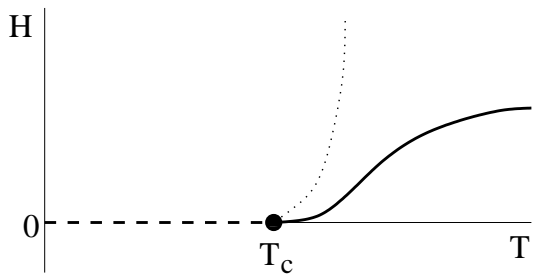}}
\caption{Phase diagrams for Ising clusters and droplets in two dimensions. The first order transition (dashed line) is common to clusters and droplets. Above $T_c$ there is a second order transition along the continuous line for clusters, and along the Kert\'esz line (dotted) for droplets.
}
\label{t-h} 
\end{figure}

In two dimensions also the Ising clusters undergo a first order transition for $H=0$, $T<T_c$ \cite{CNPR,SG}. Above $T_c$ they exhibit instead a second order transition going from $(T,H)=(T_c,0)$ to a non-negative value of $H$ at infinite temperature, where, due to the vanishing of the ferromagnetic interaction, a random percolation fixed point is located. The first and second order transition lines determine a curve in the $T$-$H$ plane above which $P>0$ (Fig.~\ref{t-h}). 

Droplets cannot percolate at infinite temperature since they have $p_B=0$ there. There is however a second order transition line, the Kert\'esz line \cite{Kertesz}, going from the Curie point to a value of $T$ at $H=+\infty$, where again a random percolation fixed point is located; $P>0$ to the left of the Kert\'esz line (Fig.~\ref{t-h}).

The Curie point $(T,H)=(T_c,0)$ is a fixed point for both Ising clusters and droplets. We consider the critical behavior of the percolative observables introduced in the previous section when this point is approached both at $H=0$ and at $T=T_c$. Denoting by $g$ the deviation from criticality, i.e. $|T-T_c|$ in the first case and $|H|$ in the second, we have for $g\to 0$
\bea
P&=&B\,g^\beta\,,
\label{beta}\\
S&=&\Gamma\,g^{-\gamma}\,,
\label{gamma}\\
\xi&=&f\,g^{-\nu}\,,
\label{nu}\\
\left(\frac{\langle N_c\rangle}{N}-\frac{M}{2}\right)_{\text{sing}}&=&\left(\frac{A_{-1}}{2}\ln^2g-A_0\ln g+\delta_{A_{-1},0}\,A_1\right)g^\mu\,.
\label{mu}
\eea
Critical exponents and critical amplitudes depend on the direction along which the critical point is approached. We distinguish the following cases:

(a)\,\, $T\to T_c^+$, $H=0$\,: we attach a subscript $a$ to amplitudes and exponents. As discussed in the next section, we will consider this limit only for droplets;

(b)\,\, $T\to T_c^-$, $H=0^\pm$\,: we attach a subscript $b$ to amplitudes and exponents and a superscript $\pm$ to the amplitudes. Droplet exponents are the same as in case (a);

(c)\,\, $T=T_c$, $H\to 0^\pm$\,: we attach a subscript $c$ to amplitudes and exponents and a superscript $\pm$ to the amplitudes.

The form of (\ref{mu}) needs to be explained. Since the approach to criticality of the Ising magnetization $M$ is known (see \cite{report}), the most interesting part of (\ref{mcn}) is the contribution coming from the dilute Potts free energy $f_q$. This is the sum of a regular part containing non-negative powers of $g$, and of a singular part,
\EQ
f_q^{\text{sing}}(g)=F_q\,g^{\mu_q}\,,\hspace{.6cm}g\to 0\,,
\label{fsing}
\EN
that we need to consider in the limit $q\to 1$. If $\mu_1$ happens to coincide with a non-negative integer, the resonance with the regular part is signalled by a pole in the amplitude $F_q$. The latter can be expanded around $q=1$ in the form
\EQ
F_q=\sum_{k=-1}^{\infty}a_k\left[\Delta\mu_q\right]^k,
\label{coeff}
\EN
with $\Delta\mu_q\equiv -(\mu_q-\mu_1)$. Evaluation of (\ref{fsing}) for $q\to 1$ then leads to
\EQ
f_\text{Ising}^\text{sing}(g)=(-a_{-1}\ln g+\delta_{a_{-1},0}\,a_0)\,g^\mu\,,\hspace{1cm}\mu\equiv\mu_1\,,
\label{fising}
\EN
while $\partial_qf_q^\text{sing}|_{q=1}$ yields (\ref{mu}) with 
\EQ
A_k=\left.\partial_{q}\mu_q\right|_{q=1}a_k\,,\hspace{1cm}k=-1,0,1\,.
\label{ak}
\EN
When $\mu$ is an integer, i.e. when $a_{-1}\neq 0$, the term coming from $A_1$ in (\ref{mu}) and that coming from $a_0$ in (\ref{fising}) contribute to the regular part. Notice that in the case of random percolation\footnote{In random percolation $g=|p-p_c|$ measures the deviation from the critical site occupation probability.} $f^\text{sing}_1(g)=0$, so that (\ref{coeff}) starts from $k=1$ and (\ref{mu}) holds with $M=A_{-1}=A_0=0$.

\section{Field theory}
As shown in \cite{isingperc}, in the {\em scaling limit} towards the Curie point, each of the two phase diagrams of Fig.~\ref{t-h} (one for clusters, one for droplets) can be seen as the $q\to 1$ projection of a phase diagram living in a three-dimensional space with coordinates $(g_1,g_2,q)$, where $g_1$ and $g_2$ are couplings which become $\tau\sim T-T_c$ and $h\sim H$, respectively, at $q=1$. This three-dimensional phase diagram is associated to the field theory with action
\EQ
{\cal A}={\cal A}_\text{CFT}-g_1\int d^2x\,\phi_1(x)-g_2\int d^2x\,\phi_2(x)\,,
\label{action}
\EN
where ${\cal A}_\text{CFT}$ is the conformal action describing the pertinent critical line (parameterized by $q$) within the scaling limit of the dilute Potts model (\ref{hamiltonian}), and $\phi_1$, $\phi_2$ are $S_q$-invariant relevant fields in this conformal theory. Recalling that two-dimensional conformal field theories \cite{BPZ} characterized by a central charge
\EQ
c=1-\frac{6}{p(p+1)}
\label{c}
\EN
contain scalar primary fields $\varphi_{r,s}$ with scaling dimension
\EQ
X_{r,s}=\frac{[(p+1)r-ps]^2-1}{2p(p+1)}\,,
\label{x}
\EN
the action (\ref{action}) that for $q\to 1$ describes the scaling limit for clusters and droplets is specified as follows \cite{isingperc}
\bea
&&\mbox{clusters:}\hspace{1.3cm}\sqrt{q}=2\sin\frac{\pi(p+2)}{2p},\hspace{1cm}\phi_1=\varphi_{1,3},\hspace{.5cm}\phi_2=\varphi_{1,2}\,;\nonumber\\
&&\mbox{droplets:}\hspace{.5cm}\sqrt{q+1}=2\sin\frac{\pi(p-1)}{2(p+1)},\hspace{1cm}\phi_1=\varphi_{2,1},\hspace{.5cm}\phi_2=\varphi_{(p-1)/2,(p+1)/2}\,.\nonumber
\eea
In the cluster case ${\cal A}_\text{CFT}$ corresponds to the tricritical line of the dilute $q$-state Potts model, and $\phi_1$, $\phi_2$ are the dilution and energy fields along this line. In the droplet case ${\cal A}_\text{CFT}$ corresponds to the critical line of the pure $(q+1)$-state Potts model (\ref{q+1}), $\phi_1$ is the energy field on this line and $\phi_2$ the spin field $\omega_0$. In both cases $q=1$ corresponds to $p=3$ and, as expected, $c=1/2$, the central charge of the critical Ising model. The Potts spin has dimension $X_s$ given by $X_{p/2,p/2}$ for clusters \cite{SV} and $X_{(p-1)/2,(p+1)/2}$ for droplets. The critical exponents in (\ref{beta})-(\ref{mu}) are given by
\EQ
\nu=\frac{1}{2-X},\hspace{.9cm}\beta=X_s|_{q=1}\nu,\hspace{.9cm}\gamma=2(1-X_s|_{q=1})\nu,\hspace{.9cm}\mu=2\nu,
\EN
with $X=1$ in cases (a) and (b), and $X=1/8$ in case (c). The exponents are collected in Table~\ref{exponents}.

\begin{table}[htbp]
\begin{center}
\begin{tabular}{|l|c|c|c|c||c|c|c|c|}
\hline
 & $\nu_b$ & $\beta_b$ & $\gamma_b$ & $\mu_b$ & $\nu_c$  & $\beta_c$ &
$\gamma_c$ & $\mu_c$ \\
\hline
clusters & $1$  & $5/96$ & $91/48$ & $2$ & $8/15$ & $1/36$ & $91/90$ & $16/15$ \\
droplets & $1$ & $1/8$ & $7/4$ & $2$ & $8/15$ & $1/15$ & $14/15$ & $16/15$ \\
\hline
\end{tabular}
\caption{\label{exponents} Critical exponents for Ising percolation.}
\label{tabcrexp}
\end{center}
\end{table}

The percolative transitions in the scaling Ising model are the $q\to 1$ limit of transitions associated to sponeneous breaking of $S_q$ symmetry in (\ref{action}). For both clusters and droplets, the first order part of the transition corresponds to $g_2=0$ and $g_1<0$. The second order part corresponds to $g_2=0$ and $g_1>0$ for clusters, while for droplets it maps onto a renormalization group trajectory with $g_1$ and $g_2$ both non-zero \cite{isingperc}. In particular, this implies that the deviation from $H=0$ of the transition above $T_c$ for clusters is entirely due to corrections to scaling, namely that the behavior associated to the limit (a) for clusters is non-universal\footnote{In particular, the second order percolative transition is expected to stay at $H=0$ for the triangular lattice, while it develops as in Fig.~\ref{t-h} for the square lattice (see \cite{isingperc}). As a consequence, the connectivity length for clusters at $T>T_c$, $H=0$ is infinite in the first case, and finite in the second.} and, in general, is not determined by the scaling theory (\ref{action}). This is why the limit (a) for clusters is excluded in our study of universal amplitude ratios.

\subsection{Integrability}
It is known from \cite{Taniguchi} that deformations of conformal field theories with $c<1$ through a single field of type $\varphi_{1,2}$, $\varphi_{2,1}$ or $\varphi_{1,3}$ are integrable. This means that cases (b),(c) for clusters and cases (a),(b) for droplets all are $q\to 1$ limits of integrable cases of (\ref{action}). Integrable field theories are solved in the $S$-matrix framework \cite{ZZ}, and we now recall the solutions for the scaling pure and dilute Potts model.

\vspace{.3cm}
{\bf$\varphi_{2,1}$ and $\varphi_{1,2}$ deformations.} 
Consider a (1+1)-dimensional integrable field theory with spontaneously broken $S_Q$ symmetry \cite{CZ}. The elementary excitations are kinks $K_{ij}(\theta)$, $i\neq j$, with mass $m$ and energy-momentum $(m\cosh\theta,m\sinh\theta)$, interpolating between pairs of degenerate vacua $|\Omega_i\rangle$, $i=1,\dots,Q$. Integrability ensures that any scattering process reduces to a sequence of elastic two-kink collisions of type
\EQ
|K_{ik}(\theta_1)K_{kj}(\theta_2)\rangle=\sum_lS_{ij}^{kl}(\theta_1-\theta_2)\,|K_{il}(\theta_2)K_{lj}(\theta_1)\rangle\,,
\label{kinkamplitudes}
\EN
where ``in'' (``out'') asymptotic states correspond to $\theta_1$ larger (smaller) than $\theta_2$. Permutational symmetry implies that there are only four different two-kink scattering amplitudes $S_{ij}^{kl}(\theta)$: indeed, there are two scattering channels ($i=j$ and $i\neq j$) and in each of them the central vacuum can preserve its color ($l=k$) or change it ($l\neq k$). The minimal solution for these amplitudes, satisfying the constraints of unitarity, crossing symmetry, factorization and bootstrap, was determined in \cite{CZ} and contains a parameter $\lambda$ which is related to $Q$ as
\EQ
\sqrt{Q}=2\sin\frac{\pi\lambda}{3}\,.
\label{lambda}
\EN
For $Q\in(0,4)$, this solution corresponds to the $\varphi_{2,1}$ deformation of the $Q$-state Potts critical line when $\lambda\in(0,3/2)$, and to the $\varphi_{1,2}$ deformation of the $Q$-state Potts tricritical line when $\lambda\in(3/2,3)$. The critical and tricritical lines meet at $Q=4$ and have central charge (\ref{c}) with
\EQ
\lambda=\left\{
\begin{array}{l}
\frac{3(p-1)}{2(p+1)}\,,\hspace{1cm}\mbox{critical line}\\ 
\\
\frac{3(p+2)}{2p}\,,\hspace{1cm}\mbox{tricritical line}\,.
\end{array}
\right.  
\EN
The spontaneously broken phase we are discussing corresponds to a specific sign of the coupling $g$ conjugated to the field responsible for the deformation; the unbroken phase corresponds to the other sign and is related to the broken phase by duality.

The full particle spectrum is determined investigating the pole structure of the amplitudes and going through the bootstrap procedure \cite{CZ,DPT}. For our purposes it is enough to know that the two lightest topologically neutral bound states $B_j$, $j=1,2$, appear for $\lambda>j$ as poles of the kink-antikink amplitudes $S_{ii}^{kl}$; they have mass $m_j=2m\sin\frac{j\pi}{2\lambda}$.

Integrability allows in particular the exact determination of the singular part of the free energy per unit area. For the $\varphi_{1,2}$ deformation it reads \cite{Fateev}
\EQ
f^\text{sing}(g,p)=-\frac{\sin\left(\frac{\pi p}{3p+6}\right)}{4\sqrt{3}\sin\left(\frac{\pi(2p+2)}{3p+6}\right)}\,m^2\,,
\label{f12}
\EN
where the kink mass is related to the coupling $g$ as
\EQ
m=\frac{2^{\frac{p+5}{3p+6}}\sqrt{3}\,\Gamma\left(\frac{1}{3}\right)\Gamma\left(\frac{p}{3p+6}\right)}{\pi\,\Gamma\left(\frac{2p+2}{3p+6}\right)}\left[\frac{\pi^2\,g^2\,\Gamma^2\left(\frac{3p+4}{4p+4}\right)\Gamma\left(\frac{1}{2}+\frac{1}{p+1}\right)}{\Gamma^2\left(\frac{p}{4p+4}\right)\Gamma\left(\frac{1}{2}-\frac{1}{p+1}\right)}\right]^{\frac{p+1}{3p+6}}\,.
\EN
Comparison with (\ref{fsing}) gives
\EQ
\mu{(p)}=\frac{4(p+1)}{3(p+2)}\,.
\label{muq}
\EN
The corresponding results for the $\varphi_{2,1}$ deformation are obtained through the replacement $p\to-p-1$ into the last three equations \cite{Fateev}.

\vspace{.3cm}
{\bf $\varphi_{1,3}$ deformation.}
Consider a $S_Q$-invariant theory on the first order transition point where the ordered vacua $|\Omega_i\rangle$, $i=1,\dots,Q$, are degenerate with the disordered vacuum $|\Omega_0\rangle$. The elementary excitations are kinks $K_{0i}$ with mass $m$, going from the disordered to the $i$-th ordered vacuum, together with their antikinks $K_{i0}$. There  are again four different two-kink amplitudes which in the notation (\ref{kinkamplitudes}) read $S_{00}^{kl}$, $S_{kl}^{00}$, where the cases $k=l$ and $k\neq l$ have to be distinguished. The minimal integrable solution was given in \cite{dilute} and corresponds to the $\varphi_{1,3}$ deformation of the $Q$-state Potts tricritical line. In this case the interaction among the kinks does not produce bound states. Again we refer to a specific sign (positive) of the coupling $g$ conjugated to $\varphi_{1,3}$, the other sign corresponding to the massless flow from the tricritical to the critical line.

For the free energy we now have \cite{Alyoshalambda-m,FLZZ}
\EQ
f^\text{sing}(g,p)=-\frac{\sin^2\frac{\pi p}{2}}{2\sin\pi p}\,m^2\,,
\label{f13}
\EN
\EQ
m=\frac{2\,\Gamma\left(\frac{p}{2}\right)}{\sqrt{\pi}\,\Gamma\left(\frac{p+1}{2}\right)}\left[\frac{\pi\,g\,(p-1)(2p-1)}{(1+p)^2}\sqrt{\frac{\Gamma\left(\frac{1}{1+p}\right)\Gamma\left(\frac{1-2p}{1+p}\right)}{\Gamma\left(\frac{p}{1+p}\right)\Gamma\left(\frac{3p}{1+p}\right)}}\right]^{\frac{1+p}{4}}\,,
\EN
\EQ
\mu(p)=\frac{p+1}{2}\,.
\EN

\subsection{Connectivity}
Within our formalism based on factorized scattering among kinks, correlators are expressed as spectral sums
\EQ
\langle\Phi(x)\Phi(0)\rangle_c=\sum_{n=1}^\infty\sum_{\gamma_1,\ldots,\gamma_{n-1}}\int_{\theta_1>\ldots>\theta_n}\frac{d\theta_1}{2\pi}\ldots\frac{d\theta_n}{2\pi}|F_{\alpha\gamma_1\ldots\gamma_{n-1}\beta}^{\Phi}(\theta_1,\ldots,\theta_n)|^2e^{-m|x|\sum_{k=1}^n\cosh\theta_k},
\label{spectral}
\EN
where the form factors
\EQ
F_{\alpha\gamma_1\ldots\gamma_{n-1}\beta}^{\Phi}(\theta_1,\ldots,\theta_n)=\langle\Omega_\alpha|\Phi(0)|K_{\alpha\gamma_1}(\theta_1)K_{\gamma_1\gamma_2}(\theta_2)\ldots K_{\gamma_{n-1}\beta}(\theta_n)\rangle
\label{ff}
\EN
can be computed exactly relying on the knowledge of the $S$-matrix (see the form factor equations in Appendix~A). The greek vacuum indices in (\ref{ff}) take the color value $k=1,\ldots,q$, and also the value $0$ when $q+1$ phases coexist; it is understood that adjacent vacuum indices cannot coincide. We included in (\ref{spectral}) only the states made of elementary kink excitations; it is understood that if there are stable bound states they also contribute to the spectral sum. It is well known that spectral series in integrable field theory converge very rapidly and that truncation of the series to the first (lightest) contribution is sufficient to provide accurate results upon integration in $d^2x$ (see in \cite{DVC} the results obtained in this way for random percolation). This is the approximation we are going to adopt also in this paper. 

It follows from (\ref{connectivity}) that the problem of determining the connectivity function $P_f(x)$ reduces to the study of the Potts connected correlator $G(x)$ in the limit $q\to 1$. This correlator vanishes at $q=1$ (no Potts degrees of freedom), and (\ref{connectivity}) shows that it vanishes linearly in $q-1$: 
\EQ
P_f(x)=\lim_{q\to 1}\frac{G(x)}{q-1}\,.
\label{pf}
\EN
The $S$-matrix does not force itself the form factors to vanish at $q=1$; the vanishing of form factors can instead be induced by the color structure of the fields and by their normalization conditions. The constraint $\sum_k\sigma_k=0$ can induce a linear vanishing of the form factors of $\sigma_1$ on some states; the contribution of these states then vanishes quadratically in the spectral decomposition of $G(x)$, and can be ignored for $q\to 1$. This means that the leading (linear) contribution in $q-1$ to $G(x)$ comes entirely from the sum over color indices in the spectral sum, i.e. from the multiplicity of form factors identified by color symmetry. 

Notice that this symmetry can identify form factors of $\sigma_1$ only through permutations of the vacuum indices $\gamma_i=2,\ldots,q$, because color $1$ is carried by the field itself\footnote{In the cases we consider the external indices $\alpha$, $\beta$ in (\ref{ff}) take values $0$ or $1$.}. It follows in particular that the states whose vacuum indices take only the values $0$ and $1$ (i.e. the states which are well defined at $q=1$ and that, for this reason, we call Ising states) cannot contribute to the multiplicity factor $q-1$, and then are among those giving a subleading contribution as $q\to 1$. Finally we conclude that the leading contribution to $G(x)$ for $|x|\to\infty$, $q\to 1$, comes from the states with minimal total mass which are not Ising states. It follows from (\ref{true}) that this minimal total mass coincides with the inverse true connectivity length $\xi_t$.

We now discuss the correlator $G(x)$, first for clusters and then for droplets, in the cases (a), (b) and (c) defined in the previous section, recalling that these are cases of (\ref{action}) with $q\to 1$: (a) (for droplets only) and (b) correspond to $g_2=0$, while (c) corresponds to $g_1=0$; the sign of $T-T_c$ and $H$ coincides with that of $g_1$ and $g_2$, respectively.

\vspace{.3cm}
{\bf Clusters.} In the case (b) we are within the $\phi_{1,3}$ deformation of the Potts tricritical line, with degenerate vacua $|\Omega_\alpha\rangle$, $\alpha=0,1,\ldots,q$. For $H=0^+$, the color symmetry is spontaneously broken and $P=\partial_q\langle\Omega_1|\sigma_1|\Omega_1\rangle|_{q=1}\neq 0$. The form factors entering the spectral sum for $G(x)$ are of type $F^{\sigma_1}_{10k0j\cdots01}$, and the lightest non-Ising contribution comes from the four-kink term $\sum_{k=2}^q|F^{\sigma_1}_{10k01}|^2=(q-1)|F^{\sigma_1}_{10201}|^2$. It follows, in particular, that $\xi_t=1/4m$. For $H=0^-$ we are in the Potts disordered vacuum and $P=\langle\Omega_0|\sigma_1|\Omega_0\rangle=0$. $G(x)$ decomposes on the form factors $F^{\sigma_1}_{0k0i\cdots j0}$ and the lightest non-Ising contribution comes from the two-kink term $\sum_{k=2}^q|F^{\sigma_1}_{0k0}|^2=(q-1)|F^{\sigma_1}_{020}|^2$; $\xi_t=1/2m$. It is interesting to compare the true connectivity length with the magnetic true correlation length $\hat{\xi}_t$ defined from the decay of the Ising spin-spin correlator,
\EQ
\langle\sigma(x)\sigma(0)\rangle_c\sim\text{e}^{-|x|/\hat{\xi}_t},\quad|x|\rightarrow\infty\,.
\label{truemagnetic}
\EN
This is now determined by the lightest Ising states in the topologically neutral sector, i.e. $\hat{\xi}_t=1/2m$ for $H=0^\pm$.

Case (c) corresponds to the $\varphi_{1,2}$ deformation of the Potts tricritical line. For $H\to 0^+$ we are in the spontaneously broken phase with degenerate vacua $|\Omega_k\rangle$, $k=1,\ldots,q$, and $P=\partial_q\langle\Omega_1|\sigma_1|\Omega_1\rangle|_{q=1}\neq 0$. It follows from what we said about this deformation and from (\ref{lambda}) that $q\to 1$ amounts to $\lambda\to 5/2$, so that the theory possesses, in particular, also the stable topologically neutral bound states $B_j$. However, the states $|B_j\rangle$ are Ising states, and the lightest non-Ising contribution to $G(x)$ comes from the term $\sum_{k=2}^q|F^{\sigma_1}_{1k1}|^2=(q-1)|F^{\sigma_1}_{121}|^2$, which implies $\xi_t=1/2m$. For $H\to 0^-$ there is instead a single, disordered vacuum, and the excitations are not kinks. This phase, however, is related to the previous one by duality \cite{DC}, so that $G(x)$ at $H\to 0^-$ coincides with $\langle\mu_j(x)\mu_j(0)\rangle$ at $H\to 0^+$, where $\mu_j(x)$ is the Potts disorder field which interpolates the kink $K_{1j}$. The lightest contribution to $G(x)$ comes then from the single one-kink term $|F^{\mu_j}_{1j}(\theta)|^2$. The latter, however, coincides \cite{DC} with $\langle\Omega_1|\sigma_1|\Omega_1\rangle|F^{\sigma_1}_{1j1}(\infty,0)|$, and then is proportional to $q-1$, as required. It also follows that $\xi_t=1/m$. Concerning the magnetic correlation lenght, the lightest topologically neutral Ising state is $|B_1\rangle$, so that $\hat{\xi}_t=1/m_1=1/(2m\sin\frac{\pi}{5})$ for $H\to 0^\pm$.

\vspace{.3cm}
{\bf Droplets.} Case (b) corresponds to the $\varphi_{2,1}$ deformation of the $(q+1)$-state Potts critical line, with $S_{q+1}$ permutational symmetry, degenerate vacua $|\Omega_\alpha\rangle$ and kinks $K_{\alpha\beta}$, $\alpha,\beta=0,1,\ldots,q$, which are the only particles for $q\leq 2$. For $H=0^+$ the $S_{q+1}$ symmetry is spontaneously broken in the direction $1$, and $\langle\Omega_1|\sigma_1|\Omega_1\rangle$ is given by the second line of (\ref{sigma1beta}). The lightest non-Ising contribution to $G(x)$ is $\sum_{k=2}^q|F^{\sigma_1}_{1k1}|^2=(q-1)|F^{\sigma_1}_{121}|^2$. For $H=0^-$ the $S_{q+1}$ symmetry is spontaneously broken in the direction $0$, so that the $S_q$ color symmetry is unbroken and $P=\langle\Omega_0|\sigma_1|\Omega_0\rangle=0$, as in the first line of (\ref{sigma1beta}). The lightest non-Ising contribution to $G(x)$ is $\sum_{k=2}^q|F^{\sigma_1}_{0k0}|^2=(q-1)|F^{\sigma_1}_{020}|^2$. We have $\xi_t=\hat{\xi}_t=1/2m$ for $H=0^\pm$.

Case (a) corresponds to the same deformation as case (b), but with the $S_{q+1}$ symmetry unbroken and a single vacuum. Relation (\ref{matching}) and use of $S_{q+1}$ invariance give $G(x)=[(q^2-1)/q^2]\langle\omega_0(x)\omega_0(0)\rangle$, which already contains the factor $q-1$. Again duality identifies $\langle\omega_0(x)\omega_0(0)\rangle$ of the unbroken phase with the correlator $\langle\Omega_1|\tilde{\omega}_j(x)\tilde{\omega}_j(0)|\Omega_1\rangle$ of the disorder field computed in the broken phase, which receives its lightest contribution from the one-kink term $|F^{\tilde{\omega_j}}_{1j}|^2$. Notice that, as in case (c) for clusters, this term can be rewritten as $\langle\Omega_1|\omega_1|\Omega_1\rangle|F^{\omega_1}_{1j1}(\infty,0)|$, but this time $\langle\Omega_1|\omega_1|\Omega_1\rangle$ does not vanish for $q\to 1$, beacause we are in a $(q+1)$-state Potts model, and this agrees with the fact that the necessary $q-1$ factor in $G(x)$ has already been obtained. For the correlation lenghts we have $\xi_t=\hat{\xi}_t=1/m$. Droplet connectivity at $H=0$ is further discussed in appendix~B.

In case (c) the theory is not integrable for $q>1$, and this eventually does not allow the computation of the form factors. We can however discuss some essential features. We deal with a $(q+1)$-state Potts model in presence of a field $\omega_0$ which explicitly breaks the symmetry down to the $S_q$ color symmetry. We see from the phase diagram of Fig.~\ref{t-h} that for $H\to 0^+$ we are inside the region with $P>0$, where the color symmetry is spontaneously broken, so that there are $q$ degenerate vacua $|\Omega_k\rangle$ and elementary kink excitations $K_{ij}$ interpolating among them. The lightest non-Ising contribution to $G(x)$ is $\sum_{k=2}^q|F^{\sigma_1}_{1k1}|^2=(q-1)|F^{\sigma_1}_{121}|^2$, which implies $\xi_t=1/2m$. For $H\to 0^-$ the color symmetry is unbroken and the vacuum is unique, but this time we are not able to use duality to make contact with the broken phase.

\section{Universal ratios}
The connections with integrable field theory discussed in the previous section allows us to compute many of the critical amplitudes defined by (\ref{beta})-(\ref{mu}), both for clusters and droplets. The amplitudes are not universal, but universal combinations can be made out of them in which metric factors cancel \cite{PHA}. 

As we saw, the ampitudes $\Gamma$ and $f$ for mean cluster size and connectivity lengths follow from the study of the Potts spin correlator $G(x)$, which determines the connectivity function $P_f(x)$. The known effectiveness of the large $|x|$ approximation, as well as the use of duality, allowed us to reduce the problem to that of the determination of some $n$-kink form factors of the Potts spin field. We saw that $n=2$ in most cases, while one case requires $n=4$. Four-kink form factors of the Potts spin for generic $q$ have not been studied in the literature, and we make no attempt to discuss them here. Concerning the two-kink form factors of the Potts spin field, complete results were obtained in \cite{dilute} for the $\varphi_{1,3}$ deformation; the $\varphi_{2,1}$ and $\varphi_{1,2}$ deformations are more complicated and only partial results are available \cite{DC,DVC}. In appendix~B we give an approximate form factor solution that we use for the evaluation of some droplets amplitudes at $H=0$.

The amplitudes $B$ of the percolative order parameter are also related to the Potts spin two-kink form factors. Indeed eq.~(\ref{iv}) of appendix~A with $n=0$ and $\Phi=\sigma_1$ relates these matrix elements to the vacuum expectation value in (\ref{mag}).

The amplitudes $A_k$ entering (\ref{mu}) follow from the $p\to 3$ limit of the free energies (\ref{f12}), (\ref{f13}), through (\ref{ak}). Phases coexisting at a first order transition point have the same free energy, as well as phases related by duality. Since $\mu=\mu_1$ is an integer in the case of the $\varphi_{1,3}$ and $\varphi_{2,1}$ deformations, $f^\text{sing}(g,p)$ has a pole at $p=3$ (i.e. $a_{-1}\neq 0$, see Table~\ref{free}), in agreement with the discussion at the end of section~3. These deformations both give the scaling Ising model with $H=0$ when $p\to 3$, and the fact that they yield the same coefficients $a_{-1}$ and $a_0$ is then expected from (\ref{fising}).

\begin{table}[htbp]
\begin{center}
\begin{tabular}{|c|r|c|c|c|}
\hline
Deformation & $a_{-1}$ & $a_0$ & $a_1$ & $ \partial_{q}\mu_q|_{q=1}$ \\
\hline
$\phi_{13}$ & $-\pi$ & $\pi(\gamma+\ln\pi)$ & - & $\frac{9}{4\pi\sqrt{3}}$\\
$\phi_{12}$ & $0$ & $-1.1977..$  & $2.7929..$  & $\frac{6}{25\pi\sqrt{3}} $ \\
$\phi_{21}$ & $-\pi$ & $\pi(\gamma+\ln\pi)$ & - & $ -\frac{4}{3\pi}$ \\
\hline
\end{tabular}
\caption{Results determining the amplitudes (\ref{ak}) for the different integrable directions. $\gamma$ is the Euler-Mascheroni constant.}
\label{free}
\end{center}
\end{table}

The results for the universal combinations of critical amplitudes that we obtain exploiting all these pieces of information are collected in Table~\ref{ratios}. They include the combinations 
\EQ
U\equiv\frac{4B^2(f_{\text{2nd}}^-)^2}{\Gamma^-}\,,\hspace{1.5cm}R\equiv A^-_0(f_{\text{t}}^-)^2\,,
\EN
whose universality follows from the scaling relations $2\beta+\gamma=2\nu=\mu$.

\begin{table}[htbp]
\begin{center}
\begin{tabular}{|l|c|c|}
\hline
 & clusters  & droplets \\
\hline
$\Gamma_a/\Gamma_b^+$ & non-universal & $40.3^{\dagger}$   \\
$f_{\text{2nd}, a}/f_{\text{t}, a}$ & '' & 0.99959..  \\
$f_{\text{t}, a}/f_{\text{t}, b}^+$ & '' & 2 \\
$f_{\text{t}, a}/\hat{f}_{\text{t}, a}$ & '' & 1 \\
$A_{k,a}/A_{k,b}^+;\text{ }k=0,-1$ & '' & 1\\
\hline
$\Gamma_b^+/\Gamma_b^-$ & - & 1 \\
$f_{\text{t},b}^+/f_{\text{t},b}^-$ & $1/2$ & 1\\
$f_{\text{2nd},b}^-/f_{\text{t},b}^-$& $0.6799$ & $0.61^{\dagger}$ \\
$f_{\text{2nd},b}^+/f_{\text{2nd}, b}^-$ & - & 1\\
$f_{\text{t},b}^+/\hat{f}_{\text{t},b}^\pm$ & $1/2$ & 1 \\
$U_b$                    &  $24.72$  & $15.2^{\dagger}$\\
$A_{k,b}^+/A_{k,b}^-;\text{ }k=0,-1$ & 1 & 1\\
$A_{0,b}^\pm/A_{-1,b}^\pm$ & $-\gamma-\ln\pi=-1.7219..$ &
$-\gamma-\ln\pi=-1.7219..$ \\
$R_b$ & $\displaystyle{\frac{3\sqrt{3}(\gamma+\ln\pi)}{64\pi^2}}=0.014165..$
& $\displaystyle{-\frac{\gamma+\ln\pi}{12\pi^2}}=-0.014539..$\\
\hline
$f_{\text{t},c}^+/f_{\text{t},c}^-$ & $1/2$ & - \\
$f_{\text{2nd},c}^-/f_{\text{t},c}^-$ & $1.002$ & - \\
$f_{\text{t},c}^+/\hat{f}_{\text{t},c}^\pm$ & $\displaystyle\sin\frac{\pi}{5}=0.58778..$ & - \\
$A_{k,c}^+/A_{k,c}^-;\text{ }k=0,1$ & 1 & - \\
$A_{0,c}^\pm/A_{1,c}^\pm$ & $-0.42883..$ & -  \\
$R_c$ & $-3.7624..\times 10^{-3}$ & - \\
\hline
\end{tabular}
\caption{Results for amplitude ratios in Ising correlated percolation. Those quoted without decimal digits or followed by dots are exact, the others are computed in the two-kink approximation; the dagger signals the use of the approximate form factor (\ref{f3approx}). Empty cases are due to ignorance of some form factors in integrable cases or, in direction (c) for droplets, to lack of integrability; ratios involving amplitudes for clusters in direction (a) are non-universal. $\gamma=0.5772..$ is the Euler-Mascheroni constant.}
\label{ratios}
\end{center}
\end{table}

In Table~\ref{ratios} the results involving only the amplitudes\footnote{We denote $\hat{f}_\text{t}$ the amplitudes of the magnetic correlation lenght defined by (\ref{truemagnetic}).} $f_\text{t}$, $\hat{f}_\text{t}$ and $A_k$ are exact. The results which involve the amplitudes $f_\text{2nd}$ and $\Gamma$, whose evaluation requires the integration of the connectivity function, are instead approximated, with the following exceptions for the droplet case. As shwon in appendix~B, droplet connectivity is the same for $H=0^\pm$, and this is why we quote that $\Gamma_b^+/\Gamma_b^-$ and $f_{\text{2nd},b}^+/f_{\text{2nd},b}^-$ are exactly equal to $1$; moreover, (\ref{magnetic}) determines the droplet connectivity in case (a) in terms of the Ising spin-spin correlator, which is exactly known \cite{WMcTB} and gives the exact result for $f_{\text{2nd},a}/f_{\text{t},a}$.

The approximated results are of two types. Those involving the truncation of the spectral series as the only approximation are expected to be very accurate, with an error that, as in other similar computations (see e.g. \cite{DVC}), is hardly expected to exceed $1\%$. Those droplet results (signalled by a dagger) which instead also rely on the use of the approximate two-kink form factor (\ref{f3approx}) could have larger errors.

We close this section discussing the issue of the correspondence between magnetic and droplet universal properties at $H=0^+$. As we saw in section~3 there is in the case an identification of the order parameters: $P=M$. This is at the origin of the fact that the magnetic correlator $\langle\sigma(x)\sigma(0)\rangle_c$ and the droplet connectivity $P_f(x)$ both diverge as $|x|^{-1/4}$ when $|x|/\xi\to 0$. In turn, this implies that the magnetic susceptibility $\chi$ and the mean droplet size $S$, which are the integrals over $x$ of these two functions, diverge with the same exponent $\gamma=7/4$ as $T\to T_c$. Equation (\ref{magnetic}) shows that the magnetic correlator actually coincides with $2P_f$ at all distances above $T_c$; the two functions, however, differ below $T_c$ due to the presence of infinite droplets. It follows that the ratio of droplet size amplitudes above and below $T_c$ does not coincide with the corresponding susceptibility ratio, a fact already pointed out in \cite{DHB}. Actually, (\ref{magnetic}) implies that the size ratio is larger than the susceptibility ratio. Our computation shows that the difference between the two ratios is not very large: our approximated result for the first, close to $40$, has to be compared with the susceptibility result $37.7$ \cite{WMcTB}. Similar remarks apply to any ratio involving integrated correlations below $T_c$.

\section{Conclusion}
In this paper we studied the universality classes of percolative critical behavior associated to clusters and droplets of ferromagnetically interacting Ising spins in two dimensions. Clusters are the connected components obtained drawing a bond between nearest neighbor positive spins; droplets are lighter objects, the bond being drawn with temperature-dependent probability $1-e^{-2/T}$. We determined universal properties of clusters and droplets in the neighborhood of the Curie point $(T,H)=(T_c,0)$, where both are critical. Remarkably, the scaling limit is integrable in zero field and, only for clusters, also along the critical isotherm. This allowed us to obtain within the $S$-matrix approach a number of results hardly accessible to perturbative field theory (the upper critical dimension is $6$ in percolative problems). In particular, we obtained lists of universal amplitude ratios for the two percolative universality classes. Many of these predictions are exact, some allow for small errors (no more than $1\%$), few of them involve an additional approximation and could be less accurate. All these results can in principle be tested through lattice numerical methods. From the theoretical point of view such a comparison would be relevant for many reasons, we mention some of them.

The field theoretical formalism leads to a picture in which percolative and magnetic observables couple to different classes of particle states. In our results this fact is immediately visible each time that the ratio between the percolative and magnetic correlation lenghts is not $1$. In this respect a particularly sharp prediction of the theory is that this ratio equals exactly $\sin(\pi/5)$ for clusters along the critical isotherm. 

The difference between percolative and magnetic properties tends to be reabsorbed, in a non-trivial way, for the droplets. Their peculiarity, indeed, is to exhibit critical exponents equal to the magnetic ones, providing in this way an alternative quantitative description of the ferromagnetic transition based on collective modes (the droplets) rather than on local spin variables. The presence below $T_c$ of infinite droplets, which contribute to the magnetic correlations but not to the connectivity within finite droplets, induces, however, differences between magnetic and percolative amplitude ratios. Our results show that these differences are not very large, but arise in the sector of the theory where our predictions could be less accurate.

The presence of a magnetic interaction among the spins is ultimately responsible for the peculiar form (\ref{mu}) of the singular part of the mean cluster number. The logarithmic terms, in particular, are absent in random percolation and, though probably challenging for lattice numerical analysis, are completely and exactly determined by the theory.

\vspace{1cm}\textbf{Acknowledgments.} Work supported in part by the ESF Grant INSTANS and by the MIUR Grant 2007JHLPEZ.

\section*{Appendix A}
The $n$-kink form factors (\ref{ff}) satisfy functional equations similar to those well known for form factors on non-topologic excitations \cite{KW,Smirnov}. For $n=2$ the kink form factor equations where considered in \cite{DC}; here we write them for any $n$:
\begin{align}
& F_{\ldots\gamma_{i-1}\gamma_{i}\gamma_{i+1}\ldots}^{\Phi}(\ldots,\theta_i,\theta_{i+1},\ldots)=\sum_{\delta}S_{\gamma_{i-1}\gamma_{i+1}}^{\gamma_i\delta}(\theta_i-\theta_{i+1})F_{\ldots\gamma_{i-1}\delta\gamma_{i+1}\ldots}^{\Phi}(\ldots,\theta_{i+1},\theta_i,\ldots)\,,
\label{i}\\
& -i\,\text{Res}_{\theta_1-\theta_2=iu_{KK}^a}F_{\alpha\gamma_1\gamma_2\ldots}(\theta_1,\theta_2,\ldots)=\nonumber\\
&\hspace{4cm}(1-\delta_{\alpha\gamma_{2}})\Gamma_{KK}^KF_{\alpha\gamma_2\ldots}^{\Phi}(\theta_a,\theta_3,\ldots)+\delta_{\alpha\gamma_2}\Gamma_{KK}^BF_{\alpha\alpha\gamma_3\ldots}^{\Phi}(\theta_a,\theta_3,\ldots)\,,
\label{ii}\\
& F_{\alpha\beta\gamma_1\ldots\gamma_{n-2}\alpha}^{\Phi}(\theta', \theta,
  \theta_1,\ldots,\theta_{n-2})=F_{\beta\gamma_1\ldots\gamma_{n-2}\alpha\beta}^{\Phi}(\theta,\theta_1,\ldots,\theta_{n-2}, \theta'-2i\pi)\,,
\label{iii}\\
&-i\,\text{Res}_{\theta'=\theta+i\pi}F^{\Phi}_{\alpha\beta\gamma_1\ldots\gamma_{n-2}\alpha}(\theta',\theta,\theta_1,\ldots,\theta_{n-2})=\delta_{\alpha\gamma_1}[F^{\Phi}_{\alpha\gamma_2\ldots\gamma_{n-2}\alpha}(\theta_1,\ldots,\theta_{n-2})\,+\label{iv}\\
&\quad-\sum_{\delta_1\ldots\delta_{n-3}}S_{\beta\gamma_2}^{\gamma_1\delta_1}(\theta-\theta_1)\ldots S_{\delta_{n-4}\gamma_{n-2}}^{\gamma_{n-3}\delta_{n-3}}(\theta-\theta_{n-3})S_{\delta_{n-3}\alpha}^{\gamma_{n-2}\beta}(\theta-\theta_{n-2})F^\Phi_{\beta\delta_1\ldots\delta_{n-3}\beta}(\theta_1,\ldots,\theta_{n-2})]\,.\nonumber
\end{align}
Equation (\ref{i}) immediately follows from the commutation relations (\ref{kinkamplitudes}). Equation (\ref{ii}) is the statement that the form factor inherits from the $S$-matrix the bound state poles corresponding to kinks ($K$) or topologically neutral particles ($B$); the residue of the scattering amplitudes on these poles determines also the three-particle couplings $\Gamma_{KK}^a$.  

\begin{figure}
\begin{center}
\includegraphics[width=2.7cm]{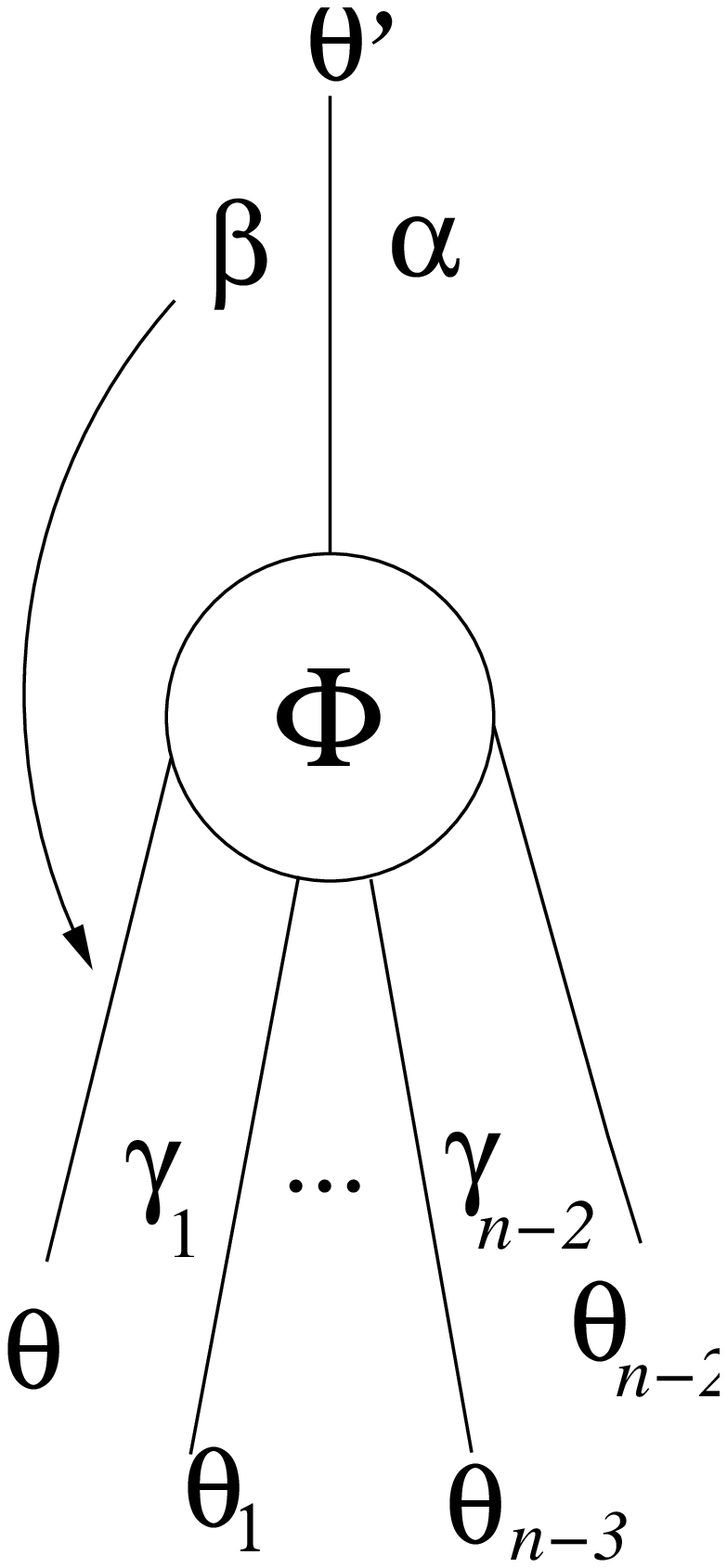}
\quad\quad
\includegraphics[width=2.5cm]{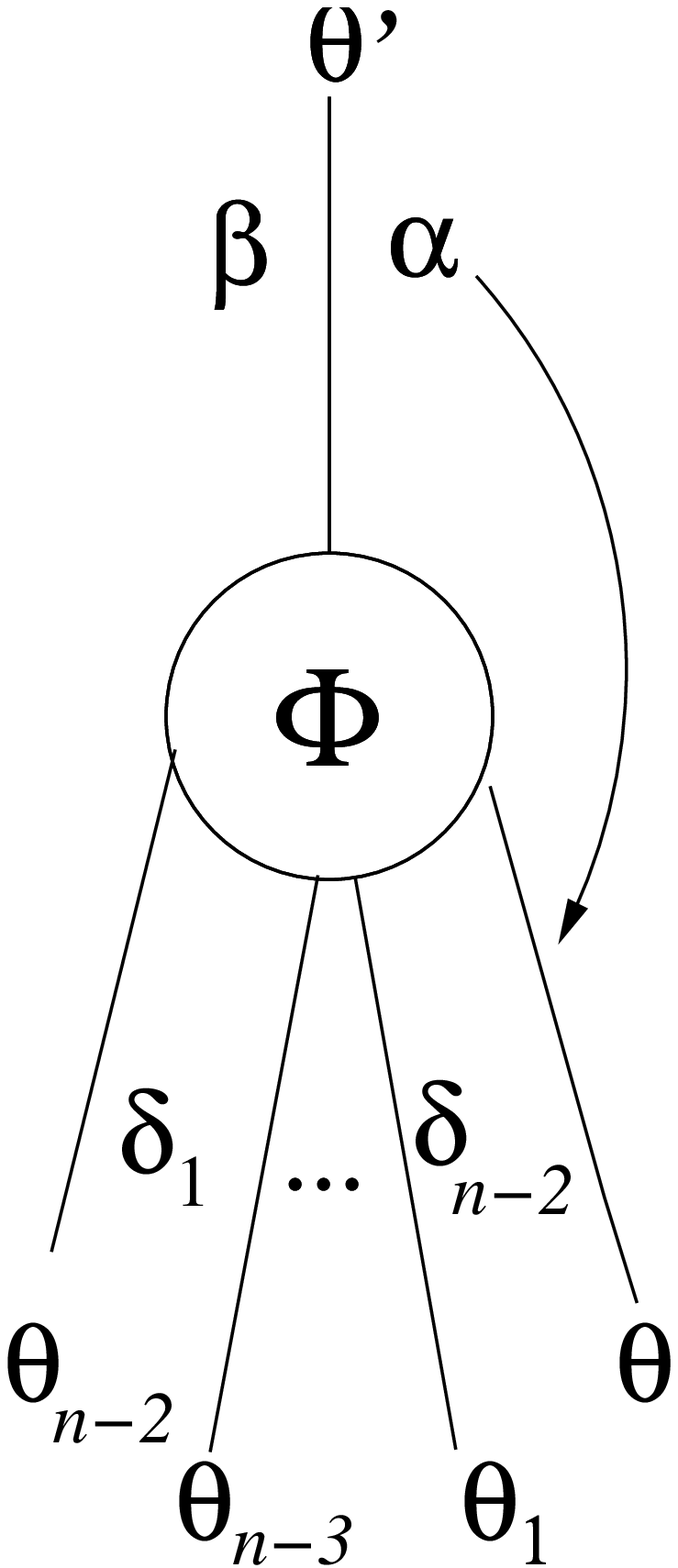}
\qquad\quad\quad
\includegraphics[width=6cm]{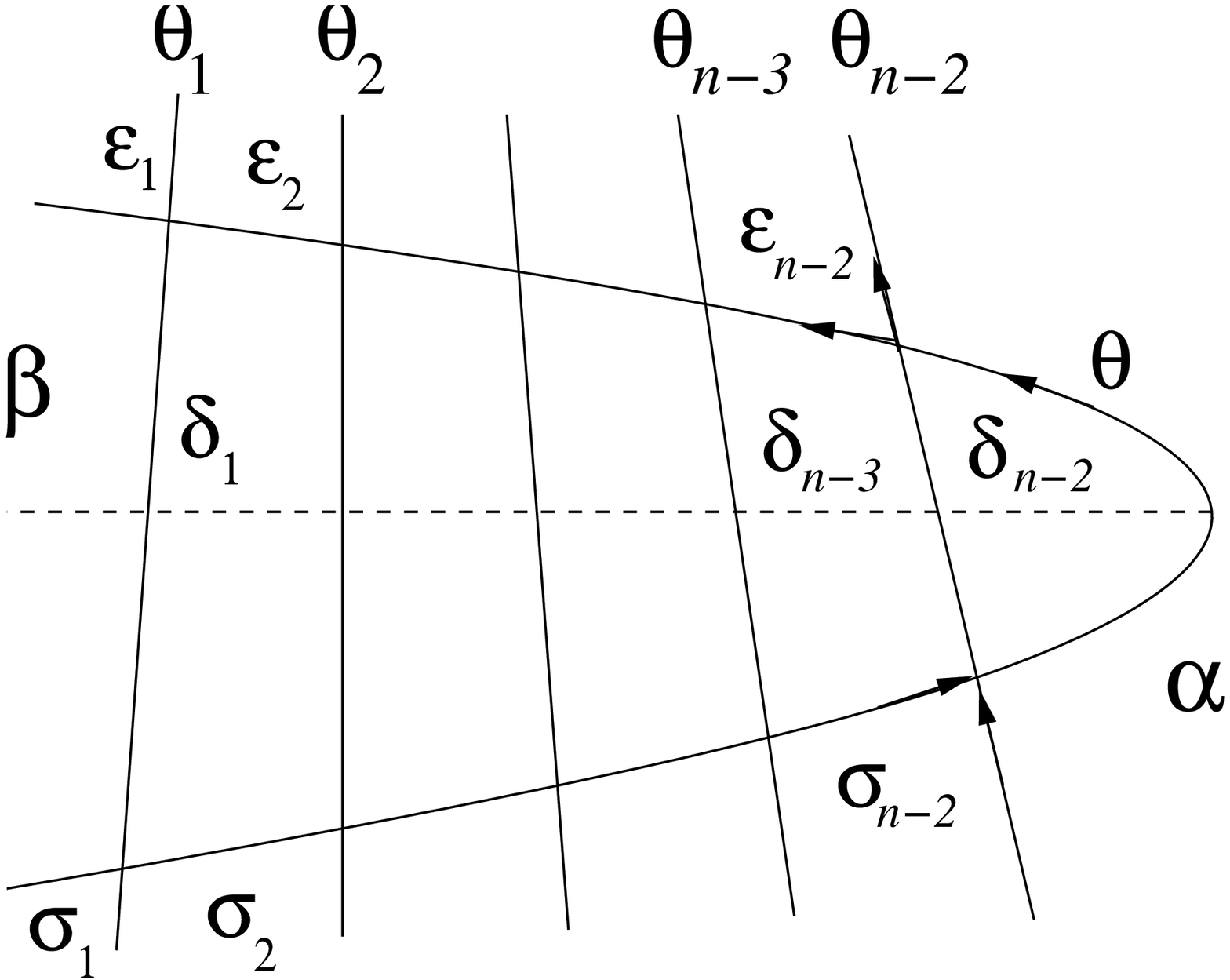}
\caption{Graphical representations of the crossing patterns in (\ref{in}) and (\ref{out}), and of the amplitude product in (\ref{inverse}).}
\label{figuraapp}
\end{center}
\end{figure}

Equations (\ref{iii}) and (\ref{iv}), that we wrote for the case of a topologically neutral field $\Phi$, can be derived adapting to the kink case an argument of \cite{Smirnov}. Consider the set of rapidities $\theta'\geq\theta>\theta_1>\ldots>\theta_{n-2}$, and recall that particles ordered with decreasing (increasing) rapidities form an ``in'' (``out'') state. The relations
\EQ
\begin{split}
\langle
&  K_{\alpha\beta}(\theta')|\Phi|K_{\beta\gamma_1}(\theta)K_{\gamma_1\gamma_2}(\theta_1)\ldots K_{\gamma_{n-2}\alpha}(\theta_{n-2})\rangle=F^{\Phi}_{\alpha\beta\gamma_1\ldots\gamma_{n-2}\alpha}(\theta'+i\pi,\theta,\theta_1,\ldots,\theta_{n-2})\\
&\hspace{7cm}+2\pi\delta(\theta'-\theta)\delta_{\alpha\gamma_1}F^{\Phi}_{\alpha\gamma_2...\gamma_{n-2}\alpha}(\theta_1,\ldots,\theta_{n-2})\,,
\end{split}
\label{in}
\EN 
\EQ
\begin{split}
\langle
& K_{\alpha\beta}(\theta')|\Phi|K_{\beta\delta_1}(\theta_{n-2})\ldots K_{\delta_{n-3}\delta_{n-2}}(\theta_1)K_{\delta_{n-2}\alpha}(\theta)\rangle=F^{\Phi}_{\beta\delta_1\ldots\delta_{n-2}\alpha\beta}(
\theta_{n-2},\ldots,\theta_{1},\theta,\theta'-i\pi)\\
&\hspace{7cm}+2\pi\delta(\theta'-\theta)\delta_{\beta\delta_{n-2}}F^{\Phi}_{\beta\delta_1\ldots\delta_{n-3}\beta}(\theta_{n-2},\ldots,\theta_1)\,,
\end{split}
\label{out}
\EN 
are pictorially shown in Fig.~\ref{figuraapp} and correspond to the crossing of the kink with rapidity $\theta'$ into an ``in'' or an ``out'' state, respectively. The term containing the delta function is a disconnected part associated to kink-antikink annihilation. We can now use (\ref{kinkamplitudes}) to reverse the ordering of the kinks with rapidities $\theta_{n-2},\ldots,\theta_1,\theta$ in (\ref{out}), with the result
\begin{multline}
\label{step1}
\sum_{\varepsilon_1...\varepsilon_{n-2}}S_{\delta_{n-3}\alpha}^{\delta_{n-2}\varepsilon_{n-2}}(\theta_{n-2}-\theta)S_{\delta_{n-4}\varepsilon_{n-2}}^{\delta_{n-3}\varepsilon_{n-3}}(\theta_{n-3}-\theta)\ldots S_{\beta\varepsilon_2}^{\delta_1\varepsilon_1}(\theta_1-\theta)\Bigl[\langle
K_{\alpha\beta}(\theta')|\Phi|K_{\beta\varepsilon_1}(\theta)...K_{\varepsilon_{n-2}\alpha}(\theta_{n-2})\rangle \Bigr.\\
-F^{\Phi}_{\beta\varepsilon_1...\varepsilon_{n-2}\alpha\beta}(\theta,\theta_1,...,\theta_{n-2}, \theta'-i\pi)\Bigl.\Bigr]=2\pi\delta(\theta'-\theta)\delta_{\beta\delta_{n-2}}F^{\Phi}_{\beta\delta_1...\delta_{n-3}\beta}(\theta_1,...,\theta_{n-2}).
\end{multline}
The relation (see Fig.~\ref{figuraapp})
\begin{equation}
\begin{split}
&\sum_{\delta_1...\delta_{n-2}} \left[S_{\beta\sigma_2}^{\sigma_1\delta_1}(\theta-\theta_1)\ldots S^{\sigma_{n-3}\delta_{n-3}}_{\delta_{n-4}\sigma_{n-2}}(\theta-\theta_{n-3}) S_{\delta_{n-3}\alpha}^{\sigma_{n-2}\delta_{n-2}}(\theta-\theta_{n-2})\,\times\right.
\label{inverse}\\
&\left.S_{\delta_{n-3}\alpha}^{\delta_{n-2}\varepsilon_{n-2}}(\theta_{n-2}-\theta)S_{\delta_{n-4}\varepsilon_{n-2}}^{\delta_{n-3}\varepsilon_{n-3}}(\theta_{n-3}-\theta)\ldots S_{\beta\varepsilon_2}^{\delta_1\varepsilon_1}(\theta_1-\theta) \right]=\delta^{\varepsilon_1}_{\sigma_1}\ldots\delta^{\varepsilon_{n-2}}_{\sigma_{n-2}}\,,
\end{split}
\end{equation}
allows to rewrite (\ref{step1}) as
\begin{multline}
\label{in2}
\langle
K_{\alpha\beta}(\theta')|\Phi|K_{\beta\gamma_1}(\theta)K_{\gamma_1\gamma_2}(\theta_1)...K_{\gamma_{n-2}\alpha}(\theta_{n-2})\rangle=F^{\Phi}_{\beta\gamma_1...\gamma_{n-2}\alpha\beta}(\theta,\theta_1,...,\theta_{n-2},\theta'-i\pi)+\\
2\pi\delta(\theta'-\theta)\sum_{\delta_1...\delta_{n-3}}S_{\beta\gamma_2}^{\gamma_1\delta_1}(\theta-\theta_1)\ldots S_{\delta_{n-3}\alpha}^{\gamma_{n-2}\beta}(\theta-\theta_{n-2})F^{\Phi}_{\beta\delta_1...\delta_{n-3}\beta}(\theta_1,...,\theta_{n-2}).
\end{multline}
Comparison of (\ref{in}) and (\ref{in2}) for $\theta\neq\theta'$ and $\theta=\theta'$ leads to (\ref{iii}) and (\ref{iv}), respectively.

\section*{Appendix B}
We saw in section~4 that the droplet connectivity in case (b) is related to $G(x)=\langle\Omega_\alpha|\sigma_1(x)\sigma_1(0)|\Omega_\alpha\rangle_c$ in the $(q+1)$-state Potts model, with $\alpha=0,1$ for $H=0^\mp$, respectively. Using (\ref{matching}), $\sum_{k=0}^q\omega_k=0$ and permutational symmetry one easily obtains
\bea
G(x)&=&\sum_{j,k=2}^q\langle\Omega_\alpha|\omega_j(x)\omega_k(0)|\Omega_\alpha\rangle_c+O((q-1)^2)\nonumber\\
&=&(q-1)[\langle\Omega_\alpha|\omega_3(x)\omega_3(0)-\omega_2(x)\omega_3(0)|\Omega_\alpha\rangle_c]+O((q-1)^2)\,,
\label{gb} 
\eea
which is the same for the two values of $\alpha$. On the other hand, (\ref{transverse}), (\ref{matching}) and permutational symmetry give for the connectivity within infinite droplets at $H=0^+$
\EQ
P_i(x)-P^2=\lim_{q\to 1}\langle\sigma_{k\neq 1}(x)\sigma_{k\neq 1}(0)\rangle_c=2\lim_{q\to 1}\langle\Omega_1|\omega_0(x)\omega_0(0)+\omega_2(x)\omega_0(0)|\Omega_1\rangle_c\,.
\label{pi}
\EN
Repeating the computation at $H=0^-$, namely on the vacuum $|\Omega_0\rangle$, gives $0$, as expected. Since for $\alpha=1$ we are free to permute $3\to 0$ in (\ref{gb}), comparison with the last equation together with (\ref{pf}) give for the magnetic correlator
\EQ
\langle\sigma(x)\sigma(0)\rangle_c=2P_f(x)+P_i(x)-P^2,\hspace{1cm}H=0^+,
\label{magnetic}
\EN
where we also used $\sigma=-2\omega_0+O(q-1)$, a consequence of (\ref{omega0}). Actually, it is easy to see computing $G(x)$ for unbroken $S_{q+1}$ symmetry that (\ref{magnetic}) holds also for $T>T_c$, where of course $P_i=P=0$.

Expanding (\ref{gb}) over kink states one recovers the result of section~4 for the droplet connectivity in case (b), namely 
\EQ
P_f(x)=\int_{\theta_1>\theta_2}\frac{d\theta_1}{2\pi}\frac{d\theta_2}{2\pi}|F(\theta_1,\theta_2)|^2e^{-m|x|(\cosh\theta_1+\cosh\theta_2)}+O(e^{-3m|x|})\,,
\label{drconn}
\EN
with
\EQ
F=F^{\sigma_1}_{\alpha 2\alpha}|_{q=1}=(F_1^{\omega}+F_3^{\omega})|_{q=1}
\label{drff}
\EN
as a consequence of (\ref{matching}) and
\EQ
F^{\omega_\alpha}_{\gamma\beta\gamma}\equiv
\delta_{\alpha\gamma}F_1^{\omega}+\delta_{\alpha\beta}F_2^{\omega}+(1-\delta_{\alpha\gamma})(1-\delta_{\alpha\beta})F_3^{\omega}\,,
\label{fj}
\EN 
\EQ
F^\omega_1+F^\omega_2+(q-1)F^\omega_3=0\,.
\EN
The form factors (\ref{fj}) were studied in \cite{DC}. For $q+1=2$, $F_1^\omega(\theta_1,\theta_2)$ is simply given by $iM_2\tanh(\theta/2)$, where $M_2=P/2$ is defined in (\ref{vev}) and $\theta=\theta_1-\theta_2$; $F_3^{\omega}(\theta_1,\theta_2)\equiv iM_2f_3(\theta)$ is the solution of the constraints\footnote{Equations (\ref{b1}), (\ref{b2}), (\ref{b3}) are the specialization of (\ref{i}), (\ref{iii}), (\ref{iv}), respectively.}
\bea
& & f_3(\theta)=-\frac{\sqrt{2}\sinh\frac{3\theta}{4}}{\sinh\left[\frac{3}{4}\left(\theta-\frac{i\pi}{3}\right)\right]}\tanh\frac{\theta}{2}+\left[\frac{\sqrt{2}\sinh\frac{3\theta}{4}}{\sinh\left[\frac{3}{4}\left(\theta-\frac{i\pi}{3}\right)\right]}-1\right]f_3(-\theta)\,,
\label{b1}\\
& & f_3(\theta+2i\pi)=f_3(-\theta)\,,
\label{b2}\\
& & \text{Res}_{\theta=i\pi}f_3(\theta)=0\,,
\label{b3}
\eea 
with the mildest asymptotic behavior as $\theta\to+\infty$.
Here we content ourselves with an approximate solution to this analytic problem. Notice first of all that (\ref{b1}) and (\ref{b2}) yield in particular $f_3(0)=0$ and $f_3(+\infty)=-i$; a solution  of (\ref{b1}) is easily checked to be $-i\tanh\frac{3\theta}{4}\tanh\frac{\theta}{2}$. If we take instead
\EQ
\tilde{f}_3(\theta)=-i\tanh\theta\tanh\frac{\theta}{2}\,,
\label{f3approx}
\EN 
we satisfy (\ref{b2}) and (\ref{b3}) at the price of badly approximating $f_3(\theta)$ near $\theta=0$, where in any case this function is vanishing and can be expected to give a small contribution to the rapidity integral in the spectral sum. The quality of the approximation is illustrated in Table \ref{f3}.

\begin{table}[htbp]
\begin{center}
\begin{tabular}{|c|c|}
\hline
rhs/lhs   & $\theta$  \\
\hline
$0.5$     & 0 \\
$0.763+0.150i$ & 1    \\
$0.933+0.061i$ &  2   \\
$0.983+0.017i$ &  3  \\
$0.996+0.004i$ & 4   \\
$0.999+0.001i$ & 5   \\
\hline
\end{tabular}
\caption{The ratio between the rhs and the lhs of (\ref{b1}) with (\ref{f3approx}) in place of $f_3$, for some values of $\theta$.}
\label{f3}
\end{center}
\end{table}


\end{document}